\documentclass[12pt]{iopart}

\usepackage[T1]{fontenc}

\usepackage{amsthm,amssymb}

\usepackage[pdftex]{graphicx}

\def\C{\mathbb C}

\def\EE{\mathcal E}

\newtheorem{lemma}{Lemma}[section]
\newtheorem{proposition}[lemma]{Proposition}

\theoremstyle{definition}

\newtheorem{remark}[lemma]{Remark}

\def\im{\mathrm i}
\def\e{\mathrm e}
\def\d{\mathrm d}
\def\arccot{\mathrm{arccot}}

\begin{document}

\title[Vertex couplings interpolating between different symmetries]{A family of quantum graph vertex couplings interpolating between different symmetries}

\author{Pavel Exner$^{1,2}$, Ond\v{r}ej Turek$^{2,3,4}$, Milo\v{s} Tater$^{1,2}$}

\address{$^1$ Doppler Institute for Mathematical Physics and Applied Mathematics, Czech Technical University in Prague, B\v{r}ehov\'{a} 7, 115 19 Prague, Czech Republic}
\address{$^2$ Department of Theoretical Physics, Nuclear Physics Institute CAS, 250 68 \v{R}e\v{z}, Czech Republic}
\address{$^3$ Department of Mathematics, Faculty of Science, University of Ostrava, 30.\ dubna 22, 701 03 Ostrava, Czech Republic}
\address{$^4$ Laboratory for Unified Quantum Devices, Kochi University of Technology, 782-8502 Kochi, Japan}
\ead{exner@ujf.cas.cz,
ondrej.turek@osu.cz,
tater@ujf.cas.cz}

\begin{abstract}
The paper discusses quantum graphs with a vertex coupling which interpolates between the common one of the $\delta$ type and a coupling introduced recently by two of the authors which exhibits a preferred orientation. Describing the interpolation family in terms of circulant matrices, we analyze the spectral and scattering property of such vertices, and investigate the band spectrum of the corresponding square lattice graph.
\end{abstract}

%
\vspace{2pc}

\section{Introduction} \label{s: intro}

Quantum graphs are useful both as a source of numerous models of nano\-structures and as a tool to examine quantum dynamics in a situation when the configuration space has a nontrivial topology \cite{BK13}. Spectral properties of quantum graphs are influenced by the conditions matching the wave function at the graphs vertices. Often the simplest conditions, usually dubbed Kirchhoff, are employed, however, there are situations when another vertex coupling may better suit the model in question; recall that \emph{any} self-adjoint coupling may be given a reasonable physical meaning \cite{EP13}.

Motivated by a recent model\footnote{Rectangular lattices were used to describe the anomalous Hall effect also in \cite{GKB09}.} of the anomalous Hall effect \cite{SK15} two of the present authors introduced and investigated in \cite{ETa18} a class of vertex couplings which are, in contrast to classes usually appearing in models, non-invariant with respect to the mirror transformation describing a sort of inherent `rotation' associated with the vertices. It was found, in particular, that the properties of such a coupling depend strongly on the vertex degree parity. From the point of view of applications, however, it is preferable to have a wider class with parameters allowing one to `tune' the vertex properties. With this goal in mind we introduce in this paper a family of couplings which interpolates between those of \cite{ETa18} and a common coupling known as $\delta$. We will examine such couplings and find their spectral and scattering properties. Our main interest is in periodic quantum graphs, thus most attention will be paid to the investigation of a square lattice with the edge length $\ell$ and the vertex coupling of the indicated type. We will analyze the spectrum of the lattice and its behavior when the vertex couplings transit from a pure $\delta$ type to the mentioned `rotational' type. The square lattice is the simplest choice and there are other cases of interest like the honeycomb one considered in \cite{ETa18}, a rectangular one, especially with incommensurate edges, etc.; they could be a subject of investigation in a later paper.

\section{Preliminaries} \label{s: prelim}

We work in the standard quantum graph framework \cite{BK13}: the state Hilbert space associated with a graph $\Gamma$ is $\bigoplus_{e\in\EE_\Gamma} L^2(e)$ where $\EE_\Gamma$ is the family of the edges of $\Gamma$, and the Hamiltonian acts as the Laplacian, $\psi_e \mapsto -\psi_e''$, where $\psi_e$ is the wave function component at the edge $e$. In other words, the motion is supposed to be free except at the graph vertices. The values of the physical constants are not important for the qualitative analysis, so we put $\hbar=2m=1$ for the sake of convenience.

To describe the vertex coupling consider a single vertex of degree $n$. The most general self-adjoint coupling is described by boundary conditions that can
be written in the matrix form \cite{GG91, Ha00}
\begin{equation}\label{bc}
(U-I)\Psi(0)+\im(U+I)\Psi'(0)=0,
\end{equation}
where $\Psi(0)=(\psi_1(0),\ldots,\psi_n(0))^T$ and
$\Psi'(0)=(\psi_1'(0),\ldots,\psi_n'(0))^T$ are the vectors of
boundary values of the wave function components and their
derivatives, conventionally taken in the outside direction, at the
vertex, and $U$ is a unitary matrix of order $n$.

Vertex couplings can be classified in various ways, in particular, several important families of vertex couplings can be distinguished by symmetries. A symmetry is typically described by an invertible map in the space of the boundary values, $\Theta:\,\C^n\to\C^n$, or a family of such maps. A vertex coupling is symmetric with respect to $\Theta$ if the condition (\ref{bc}) is equivalent to
\begin{equation*}
(U-I)\Theta\Psi(0)+\im(U+I)\Theta\Psi'(0)=0\,,
\end{equation*}
or in other words, if $U$ obeys the identity
\begin{equation}\label{sym}
\Theta^{-1}U\Theta=U.
\end{equation}
Let us mention some commonly used symmetries.
\begin{itemize}
\setlength{\itemsep}{-1.5pt}
\item \emph{Mirror symmetric} couplings have the conditions (\ref{bc}) invariant against the replacement $(\psi_1(0),\ldots,\psi_n(0))\mapsto(\psi_n(0),\ldots,\psi_1(0))$ and the same for the derivatives; in other words, $\Theta=A$, the anti-diagonal matrix with the entries equal to $1$ on the main antidiagonal and zero otherwise.
\item \emph{Permutation-invariant} couplings are a subset of the above class. In this case the conditions (\ref{bc}) are invariant with respect to any simultaneous permutation of the entries of $\Psi(0)$ and $\Psi'(0)$; in other words, the matrices $\Theta$ here form a representation of the symmetry group $S_n$. It is not difficult to check \cite{ET04} that for any $n$ there is just a two-parameter family of such couplings, $U=aI+bJ$, where $I$ is the unit matrix and $J$ denotes the matrix with all the entries equal to one; the parameters have to fulfil the conditions $|a|=1$ and $|a+nb|=1$.
\item \emph{Time-reversal-invariant} couplings: in this case $\Theta$ appearing in (\ref{sym}) is the antilinear operator describing the switch in the time direction. In our simple model $\Theta$ is just the complex conjugation, and using relations $U^t\bar{U}=\bar{U}U^t=I$ we find easily that the matrix describing the coupling must be now invariant with respect to the transposition, $U=U^t$.
\end{itemize}

A class important in this paper are \emph{rotationally symmetric} vertex coupling for which the condition (\ref{bc}) are independent of cyclic permutations of the entries of $\Psi(0)$ and $\Psi'(0)$. This corresponds to the map $\Theta$ of the following form,
\begin{equation}\label{P}
\Theta=R:=\left(\begin{array}{cccccc}
0 & 1 & 0 & \cdots & 0 & 0 \\
0 & 0 & 1 & \cdots & 0 & 0 \\
\vdots & \vdots &  & \ddots &  & \vdots \\
0 & 0 & 0 &  & 1 & 0 \\
0 & 0 & 0 & \cdots & 0 & 1 \\
1 & 0 & 0 & \cdots & 0 & 0
\end{array}\right)\,.
\end{equation}
Note that while the classes of mirror symmetric and time-reversal-invariant couplings do not coincide in general, their intersections with the rotationally symmetric class do, in other words, we can make the following claim:
\begin{proposition}\label{Prop.symmetries}
A rotationally symmetric vertex coupling is mirror symmetric if and only if it is time-reversal-invariant.
\end{proposition}
\noindent Note also that while the notions of permutation-invariant and time-reversal-invariant couplings are universal in the sense that they do not require embedding in an ambient space, the other two mentioned above make sense only if we think of the graph $\Gamma$ as of embedded in a plane.

It is obvious that once the condition (\ref{sym}) holds for $\Theta=R$, then it holds for $\Theta=R^j,\: j=2,3,\ldots,n-1$, as well. The matrices $U$
satisfying this requirement are a subset in the family of \emph{circulant matrices}, which generally take the form
\begin{equation}\label{C}
\left(\begin{array}{ccccc}
c_{0} & c_{1} & \cdots & c_{n-2} & c_{n-1} \\
c_{n-1} & c_{0} & c_{1} &  & c_{n-2} \\
\vdots & c_{n-1} & c_{0} & \ddots & \vdots \\
c_{2} &  & \ddots & \ddots & c_{1} \\
c_{1} & c_{2} & \cdots & c_{n-1} & c_{0}
\end{array}\right)\,,
\end{equation}
being a particular case of Toeplitz matrices. A circulant matrix is fully determined by its first row $c=(c_0,c_1,\ldots,c_{n-1})$
which is called \emph{generator} of $C$.
Let us mention that unitary circulant matrices have various applications in quantum physics, including the most common one in a treatment of mutually unbiased bases~\cite{Iv81, Go13}.

The notion of a circulant matrix is crucial to tackle our problem. Let us recall basic properties of such matrices, which will be important in the sequel. For any circulant matrix given by (\ref{C}), the vectors
\begin{equation*}
v_k=\frac{1}{\sqrt{n}}\left(1,\omega^k,\omega^{2k},\ldots,\omega^{(n-1)k}\right)^T\,,
\quad k=0,1,\ldots,n-1\,,
\end{equation*}
where $\omega:=\mathrm{e}^{2\pi\im/n}$, are its normalized eigenvectors; the corresponding eigenvalues are
\begin{equation}\label{eigenvalues}
\lambda_k=c_0+c_{1}\omega^k+c_{2}\omega^{2k}+\cdots+c_{n-1}\omega^{(n-1)k}\,,
\quad k=0,1,\ldots,n-1\,.
\end{equation}
Every circulant matrix $C$ is diagonalized using the Discrete Fourier Transform (DFT) matrix
\begin{equation*}
F=\left(\begin{array}{cccccc}
1 & 1 & 1 & 1 & \ldots & 1 \\
1 & \omega & \omega^2 & \omega^3 & \ldots & \omega^{(n-1)} \\
1 & \omega^2 & \omega^4 & \omega^6 & \ldots & \omega^{2(n-1)} \\
\vdots & \vdots & \vdots & \vdots &  & \vdots \\
1 & \omega^{n-1} & \omega^{2(n-1)} & \omega^{3(n-1)} & \ldots & \omega^{(n-1)^2} \\
\end{array}\right)\,.
\end{equation*}
In other words,
\begin{equation}\label{Fourier}
D=\frac{1}{n}F^*CF
\end{equation}
is a diagonal matrix with $\lambda_0,\lambda_1,\ldots,\lambda_{n-1}$ on the diagonal. Since $F^{-1}=\frac{1}{n}F^*$,
(\ref{Fourier}) can be rewritten as
\begin{equation}\label{DFT}
C=\frac{1}{n}FDF^*\,,
\end{equation}
which allows one to express the values $c_j$, the entries of the generator of $C$, in terms of $\lambda_k$ as follows
\begin{equation}\label{invDFT}
c_j=\frac{1}{n}\left(\lambda_0+\lambda_1\omega^{-j}+\lambda_2\omega^{-2j}+\cdots+\lambda_{n-1}\omega^{-(n-1)j}\right)\,.
\end{equation}
A circulant matrix is unitary if and only if its eigenvalues satisfy $|\lambda_j|=1$ for all $j=0,1,\ldots,n-1$. Consequently, the family of circulant unitary matrices of order $n$ can be parametrized by an $n$-tuple of real numbers from the interval $[0,2\pi)$ that represent the arguments of the eigenvalues.

\section{A class of interpolating couplings}\label{Sect.Coupling}

A prominent example of the vertex coupling which belongs to all the symmetry classes described in Section~\ref{s: prelim} is the $\delta$ coupling given by conditions (\ref{bc}) with a unitary matrix
\begin{equation}\label{U0}
U=-I+\frac{2}{n+\im\alpha}J\,,
\end{equation}
where $\alpha\in\mathbb{R}$ is the parameter of the coupling. The $\delta$ coupling is usually interpreted as a point potential of strength $\alpha$ at the vertex. Let us remark that it is well known (and can be easily demonstrated) that $\delta$ coupling is the only coupling with the wave functions continuously matched. The particular case $\alpha=0$ is known under different names, most often it is referred to as the Kirchhoff coupling.

While the $\delta$ couplings have all the indicated symmetries, by contrast the `extremal' rotationally symmetric coupling \cite{ETa18} associated with the choice
\begin{equation}\label{U1}
U=R
\end{equation}
with $R$ defined in (\ref{P}) does not have the mirror symmetry. This fact also means that the corresponding dynamics is not time-reversal-invariant as indicated in Proposition~\ref{Prop.symmetries}.

The main aim of the present paper is to introduce and examine a continuous parametric family of hybrid-type rotationally symmetric couplings that interpolate between the mirror-symmetric $\delta$ coupling, associated with the unitary matrix~(\ref{U0}), and the mirror-asymmetric `extremal' coupling, associated with the unitary matrix~(\ref{U1}). Both matrices~(\ref{U0}) and~(\ref{U1}) are circulant and so should be the sought unitary matrices representing the hybrid couplings. We are thus looking for a family of unitary matrices $\{U(t):\:t\in[0,1]\}$ such that
\begin{equation}\label{requirements on U(t)}
\renewcommand{\arraystretch}{1.2}
\left.\begin{array}{l}
U(0)=-I+\frac{2}{n+\im\alpha}J \quad\mbox{and}\quad U(1)=R; \\
\mbox{the map } t\mapsto U(t) \mbox{ is continuous on } [0,1]; \\
U(t) \mbox{ is unitary circulant for all } t\in[0,1].
\end{array}\quad\right\}
\end{equation}
To achieve this goal, we will take advantage of the properties of circulant matrices recalled in Section~\ref{s: prelim}. In particular, we will employ the relation between the matrix entries and the eigenvalues expressed in formul{\ae} (\ref{eigenvalues}) and (\ref{invDFT}). Our strategy is
to find a vector function $\Lambda(t)=(\lambda_0(t),\lambda_1(t), \ldots,\lambda_{n-1}(t))$ defined on $t\in[0,1]$ that has the following
properties:
\begin{itemize}
\item[(i)] $\Lambda(0)$ is a vector of eigenvalues of $U(0)$, obtained by (\ref{eigenvalues});
\item[(ii)] $\Lambda(1)$ is a vector of eigenvalues of $U(1)$, obtained by (\ref{eigenvalues});
\item[(iii)] $\Lambda(t):[0,1]\to\mathbb{C}^n$ is continuous;
\item[(iv)] $|\lambda_k(t)|=1$ holds for all $t\in[0,1]$ and $k\in\{0,1,\ldots,n-1\}$.
\end{itemize}
Once we find a $\Lambda(t)$ with properties (i)--(iv), then it is straightforward to use formula~(\ref{invDFT}) to compute the generators of circulant matrices associated to each $\Lambda(t)$. Property (iv) guarantees that the circulant matrices are unitary for each $t\in[0,1]$, and property (iii) ensures that the matrices $U(t)$ depend continuously on $t$.

Following the outlined strategy, let us at first find the eigenvalues of matrices $U(0)$ and $U(1)$ given by (\ref{U0}) and (\ref{P}), respectively, using formula (\ref{eigenvalues}). For $U(0)$, we obtain
\begin{equation}\label{ev0}
\lambda_k=-1+\frac{2}{n+\im\alpha}\sum_{j=0}^{n-1}\omega^{jk}=
\left\{\begin{array}{ll}
\frac{n-\im\alpha}{n+\im\alpha} & \mbox{for } k=0\,; \\[.5em]
-1 & \mbox{for } k\geq1\,.
\end{array}\right.
\end{equation}
For the sake of brevity, let us set
\begin{equation}\label{gamma}
\gamma:=\arg\frac{n+\im\alpha}{n-\im\alpha}\in(-\pi,\pi)\,;
\end{equation}
i.e., $\frac{n-\im\alpha}{n+\im\alpha}=\e^{-\im\gamma}$. Note that one has
\begin{equation*}
\gamma\:\left\{\begin{array}{ll}
=0 & \mbox{for } \alpha=0; \\[.3em]
\in(0,\pi) & \mbox{for } \alpha>0; \\[.3em]
\in(-\pi,0) & \mbox{for } \alpha<0.
\end{array}
\right.
\end{equation*}
Similarly, applying formula (\ref{eigenvalues}) for the circulant matrix $U(1)$, we obtain
\begin{equation}\label{ev1}
\lambda_k=\omega^k \qquad \mbox{for } \,k=0,1,\ldots,n-1.
\end{equation}
With (\ref{ev0}) and (\ref{ev1}) in mind, we can construct the vector function $\Lambda(k)$ as follows:
\begin{equation}\label{ev t}
\lambda_k(t)=
\left\{\begin{array}{cl}
\e^{-\im(1-t)\gamma} & \mbox{for } k=0; \\[.5em]
-\e^{\im\pi t\left(\frac{2k}{n}-1\right)}  & \mbox{for } k\geq1
\end{array}\right.
\end{equation}
for all $t\in[0,1]$. These $\lambda_k(t)$ are obviously continuous functions of $t$ such that $\lambda_k(0)$ coincides with (\ref{ev0}), $\lambda_k(1)$ coincides with (\ref{ev1}), and $|\lambda_k(t)|=1$ holds for all $k=0,1,\ldots,n-1$. In other words, $\Lambda(t) =(\lambda_0(k),\ldots, \lambda_{n-1}(k))$ given by (\ref{ev t}) has the required properties (i)--(iv). It remains to express the entries of the sought circulant unitary matrices $U(t)$ using formula (\ref{invDFT}). This gives
\begin{equation}\label{c U(t)}
c_j(t)=\frac{1}{n}\left(\e^{-\im(1-t)\gamma}-\sum_{k=1}^{n-1}\e^{\im\pi
t\left(\frac{2k}{n}-1\right)}\cdot\omega^{-kj}\right)\,.
\end{equation}
If $t\in(0,1)$, formula (\ref{c U(t)}) can be simplified to
\begin{equation*}
c_j(t)=\frac{1}{n}\left(\e^{-\im(1-t)\gamma}+\e^{-\im\pi
t}-\frac{\e^{\im\pi t}-\e^{-\im\pi
t}}{\e^{\frac{2\im\pi}{n}(t-j)}-1}\right),
\end{equation*}
while for $t=0$ and $t=1$, we get the generators of matrices $U(0)=-I+\frac{2}{n+\im\alpha}J$ and $U(1)=R$, respectively.
The parametric family of circulant matrices $U(t)$ with generator $(c_0(t),c_1(t), \ldots,c_{n-1}(t))$ given by (\ref{c U(t)}) satisfies the requirements (\ref{requirements on U(t)}) and represents the main result of this section.

\section{Spectrum of a star graph} \label{s:star}

If we want to understand how the vertex matching conditions influence  the properties of a quantum graph Hamiltonian, it is natural to start with a graph having a single vertex.  Let thus $\Gamma$ be the star graph with $n$ semi-infinite edges and suppose that the boundary values at the vertex are matched through conditions~(\ref{bc}) with $U(t)$ defined in Section~\ref{Sect.Coupling}. Since the resolvent of the corresponding Hamiltonian differs from that of $n$ disconnected halflines by a finite rank operator, the essential/continuous spectrum of this system is the interval $[0,\infty)$. Our topic in this section are isolated negative eigenvalues, in particular, we are going to show that the negative spectrum is nonempty for any $t\in(0,1]$ and $n\geq3$. Let us write the wave function components on the graph edges as $\psi_j(x)= b_j\e^{-\kappa x}$, where $-\kappa^2$ is the sought eigenvalue. Plugging this Ansatz into (\ref{bc}), we get a system of equations for the coefficients $b_j$ which has a nontrivial solution if and only if
\begin{equation*}
\det[(U(t)-I)-\im\kappa(U(t)+I)]=0\,,
\end{equation*}
and this is further equivalent to
\begin{equation*}
\kappa=-\im\,\frac{\lambda_j(t)-1}{\lambda_j(t)+1}
\end{equation*}
for some $j\in\{0,1,\ldots,n-1\}$, where $\lambda_j(t)$ are the eigenvalues of $U(t)$ defined in (\ref{ev t}). Hence a $\kappa>0$ solves our problem in the following cases:
\begin{equation}\label{kappa 0}
\kappa=-\tan\frac{(1-t)\gamma}{2} \qquad \mbox{for $\alpha<0$ and $t\neq1$}
\end{equation}
(where $\gamma$ stands for $\arg\frac{n+\im\alpha}{n-\im\alpha}$,
cf.~(\ref{gamma})), and
\begin{equation}\label{kappa 1}
\kappa=-\cot\left(\frac{j}{n}-\frac{1}{2}\right)\pi t \qquad \mbox{for $n\geq3$ and $j<\frac{n}{2}$}\,.
\end{equation}
These solutions in turn give rise to the eigenvalues $-\kappa^2<0$ of the star graph Hamiltonian:
\begin{itemize}
\item there is a negative eigenvalue $-\tan^2\frac{(1-t)\gamma}{2}$ whenever $\alpha<0$;
\item if $n\geq3$, there is an additional $\lfloor\frac{n-1}{2}\rfloor$-tuple of eigenvalues for every $t\in(0,1]$, which take the form $-\cot^2\left(\frac{j}{n}-\frac{1}{2}\right)\pi t$ with $j$ running through $1,\ldots,\frac{n-1}{2}$ for $n$ odd and $1,\ldots,\frac{n}{2}-1$ for $n$ even.
\end{itemize}
Finally, let us comment on the behavior of the negative eigenvalues corresponding to (\ref{kappa 0}) and (\ref{kappa 1}) in the limits $t\to0+\,$ ($\delta$ coupling) and $t\to1-$ (the `extremal' rotational coupling):
\begin{itemize}
\item If $t\to0+$, all the eigenvalues $-\kappa^2$ diverge to $-\infty$ except for $-\tan^2\frac{(1-t)\gamma}{2}$ occurring for $\alpha<0$, which approaches $-\tan^2(\gamma/2)=-\alpha^2/n^2$ (note that $-\tan^2(\gamma/2)=-\alpha^2/n^2$ holds by~(\ref{gamma})). This is in accordance with the known fact that for $t=0$ the system has only one simple negative eigenvalue $-\alpha^2/n^2$ if $\alpha<0$, while for $\alpha\geq0$ its negative spectrum is empty.

\item If $t\to1-$, the eigenvalues $-\kappa^2$ approach $0$ and $-\tan^2\frac{j}{n}$, respectively. When $t=1$, the only negative eigenvalues are $-\tan^2\frac{j}{n}$ with $j$ taking values $1,\ldots,\frac{n-1}{2}$ for $n$ odd and $1,\ldots,\frac{n}{2}-1$ for $n$ even -- cf.~\cite{ETa18}.
\end{itemize}

\section{On-shell S-matrix} \label{s:smatrix}

The on-shell S-matrix corresponding to the vertex coupling associated with a unitary matrix $U$ appearing in the boundary conditions (\ref{bc}) is generally given by the formula
\begin{equation}\label{S(k)}
\mathcal{S}(k)=\left((k+1)I+(k-1)U\right)^{-1}\left((k-1)I+(k+1)U\right),
\end{equation}
where $k=\sqrt{E}$ is the momentum \cite[Sec.~2.1]{BK13}. If the coupling is rotationally invariant, the matrix $U$ is circulant, and thus obeys
\begin{equation}\label{U circ}
U=\frac{1}{n}FDF^*
\end{equation}
for $D$ being a diagonal matrix with the eigenvalues $\lambda_0,\lambda_1,\ldots,\lambda_{n-1}$ of $U$ on the diagonal, cf.~(\ref{DFT}). Plugging~(\ref{U circ}) into~(\ref{S(k)}), we obtain the on-shell S-matrix of a general rotationally invariant vertex coupling:
\begin{equation}\label{S circ}
\mathcal{S}(k)=\frac{1}{n}F\left((k+1)I+(k-1)D\right)^{-1}\left((k-1)I+(k+1)D\right)F^*\,.
\end{equation}
It is a circulant matrix as well; its eigenvalues are related to the eigenvalues of $U$ by the formula
\begin{equation}\label{ev circ}
\mu_j=\frac{k-1+(k+1)\lambda_j}{k+1+(k-1)\lambda_j} \qquad \mbox{for } j=0,1,\ldots,n-1.
\end{equation}
Formul{\ae} (\ref{S circ}) and (\ref{ev circ}) represent an easy method to find the S-matrix and its spectrum for any rotationally symmetric vertex coupling described by a unitary matrix with eigenvalues $\lambda_0,\lambda_1,\ldots,\lambda_{n-1}$.

Let us apply this approach to examine the properties of the S-matrix for the hybrid coupling with eigenvalues (\ref{ev t}) which we introduced in Section~\ref{Sect.Coupling}. Consider first the asymptotics of $\mathcal{S}(k)$ at $k\to\infty$. Equation~(\ref{ev circ}) implies
\begin{equation*}
\lim_{k\to\infty}\mu_j=\left\{\begin{array}{cl}
1 & \mbox{for } \lambda_j\neq-1; \\[.5em]
-1  & \mbox{for } \lambda_j=-1.
\end{array}\right.
\end{equation*}
At the same time, in view of formula (\ref{ev t}) (and taking into account that $|(1-t)\gamma|<\pi$) we have
\begin{equation*}
\lambda_j(t)=-1 \quad\Leftrightarrow\quad j=\frac{n}{2}\,.
\end{equation*}
Combining these two facts, we get the following result:
\begin{itemize}
\item If $n$ is odd, then $\lim_{k\to\infty}\mu_j=1$ for all $j=0,1,\ldots,n-1$. Formula (\ref{invDFT}) applied to $\mu_0,\mu_1,\ldots,\mu_{n-1}$ gives
\begin{equation*}
\lim_{k\to\infty}\mathcal{S}(k)=I.
\end{equation*}
\item If $n$ is even, then $\lim_{k\to\infty}\mu_j=1$ for $j\neq\frac{n}{2}$, while $\lim_{k\to\infty}\mu_{n/2}=-1$. Applying formula (\ref{invDFT}), one finds the generator of $\lim_{k\to\infty}\mathcal{S}(k)$ in form
\begin{equation}\label{S(infty) n even}
\left(1-\frac{2}{n},\frac{2}{n},-\frac{2}{n},\frac{2}{n},\ldots,-\frac{2}{n},\frac{2}{n}\right).
\end{equation}
\end{itemize}
Consequently, the S-matrix corresponding to the hybrid coupling from Section~\ref{Sect.Coupling} behaves at high energies differently for odd and even $n$, similarly as in the particular case $t=1$ discussed in \cite{ETa18}.

Now we will proceed to the explicit construction of $\mathcal{S}(k)$. Having in mind the question to be addressed in Section~\ref{Sect.square}, let us find the on-shell S-matrix corresponding to the coupling introduced in Section~\ref{Sect.Coupling} in the particular case $n=4$. Formula (\ref{ev t}) gives the eigenvalues of the unitary matrix $U$ as follows,
$$
\lambda_0(t)=\e^{-\im\gamma(1-t)}\;;\quad \lambda_1(t)=-\e^{-\im\pi t/2}\;;\quad \lambda_2(t)=-1\;;\quad \lambda_3(t)=-\e^{\im\pi t/2}\;.
$$
Therefore, in view of (\ref{ev circ}), the eigenvalues of $\mathcal{S}(k)$ are
$$
\mu_0=\frac{k-\im\tan\frac{\gamma(1-t)}{2}}{k+\im\tan\frac{\gamma(1-t)}{2}}\;;\quad \mu_1=\frac{\im k\tan\frac{\pi t}{4}-1}{\im k\tan\frac{\pi t}{4}+1}\;;\quad \mu_2=-1\;;\quad \mu_3=\frac{\im k\tan\frac{\pi t}{4}+1}{\im k\tan\frac{\pi t}{4}-1}\;.
$$
Using the inverse DFT formula (\ref{invDFT}) with eigenvalues $\mu_j$, one obtains the entries of the generator of $\mathcal{S}(k)$ in the following form,
\begin{eqnarray*}
[\mathcal{S}(k)]_{00}&=&\frac{1}{2\left(1+\frac{\im}{k}\tan\frac{(1-t)\gamma}{2}\right)}-\frac{1}{1+\left(k\tan\frac{\pi t}{4}\right)^2}\;; \\
{}[\mathcal{S}(k)]_{01}&=&\frac{1}{2\left(1+\frac{\im}{k}\tan\frac{(1-t)\gamma}{2}\right)}+\frac{k\tan\frac{\pi t}{4}}{1+\left(k\tan\frac{\pi t}{4}\right)^2}\;; \\
{}[\mathcal{S}(k)]_{02}&=&-1+\frac{1}{2\left(1+\frac{\im}{k}\tan\frac{(1-t)\gamma}{2}\right)}+\frac{1}{1+\left(k\tan\frac{\pi t}{4}\right)^2}\;; \\
{}[\mathcal{S}(k)]_{03}&=&\frac{1}{2\left(1+\frac{\im}{k}\tan\frac{(1-t)\gamma}{2}\right)}-\frac{k\tan\frac{\pi t}{4}}{1+\left(k\tan\frac{\pi t}{4}\right)^2}\;.
\end{eqnarray*}
Notice that at high energies we get $\lim_{k\to\infty}([\mathcal{S}(k)]_{00},[\mathcal{S}(k)]_{01},[\mathcal{S}(k)]_{02},[\mathcal{S}(k)]_{03}) =\left(\frac{1}{2},\frac{1}{2}, -\frac{1}{2},\frac{1}{2}\right)$, in agreement with (\ref{S(infty) n even}).

\section{Square lattice}\label{Sect.square}

\begin{figure}[h]
\begin{center}
\includegraphics[clip, trim=1cm 11.5cm 5cm 11.5cm,width=1\textwidth]{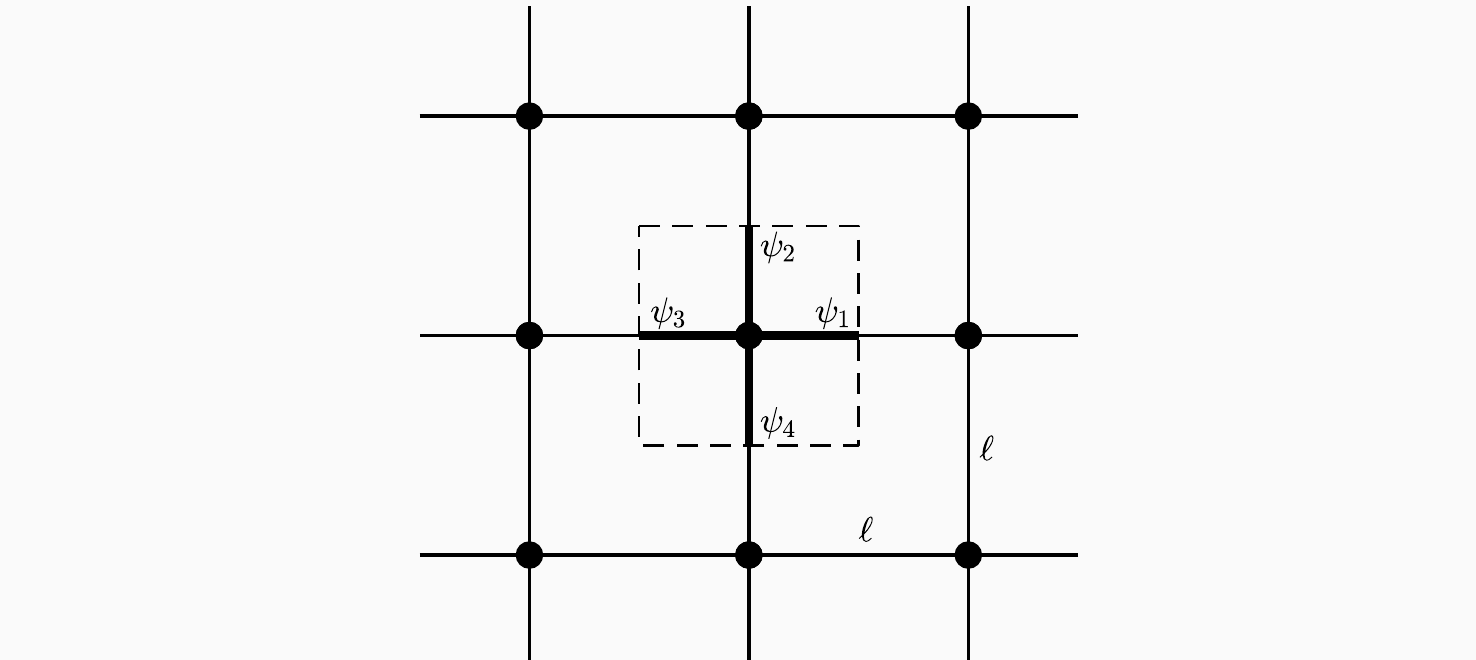}
\caption{Square lattice}\label{Mrizka}
\end{center}
\end{figure}

Consider now a square lattice with the edge length $\ell$ and boundary conditions associated with circulant unitary matrices $U(t)$, $t\in[0,1]$, defined in (\ref{c U(t)}). The whole system is symmetric with respect to rotations by integer multiples of $\pi/2$ due to the circulant character of $U(t)$. Our aim is to determine its spectrum; in view of the periodicity the natural tool to use is Bloch--Floquet decomposition. The elementary cell is depicted in Figure~\ref{Mrizka}.

Since the Hamiltonian acts as $\psi(x)\mapsto-\psi''(x)$ on each component of the wavefunction, the generalized eigenfunction Ansatz on each edge is a linear combination of $\e^{\im kx}$ and $\e^{-\im kx}$, in other words
\begin{equation}\label{vlnfce}
\eqalign{
\psi_1(x)&=C_1^+\e^{\im k x}+C_1^-\e^{-\im k x}\,,\quad x\in[0,\ell/2]\,;\\
\psi_2(x)&=C_2^+\e^{\im k x}+C_2^-\e^{-\im k x}\,,\quad x\in[0,\ell/2]\,;\\
\psi_3(x)&=C_3^+\e^{\im k x}+C_3^-\e^{-\im k x}\,,\quad x\in[-\ell/2,0]\,;\\
\psi_4(x)&=C_4^+\e^{\im k x}+C_4^-\e^{-\im k x}\,,\quad x\in[-\ell/2,0]\,.
}
\end{equation}
The boundary conditions imposed at the vertex take the form
\begin{equation}\label{vazba}
(U(t)-I)\left(\begin{array}{c}\psi_1(0)\\ \psi_2(0)\\ \psi_3(0)\\ \psi_4(0)\end{array}\right)+\im(U(t)+I)\left(\begin{array}{c}\psi_1'(0)\\ \psi_2'(0)\\ -\psi_3'(0)\\ -\psi_4'(0)\end{array}\right)=0\,.
\end{equation}
In accordance with Figure~\ref{Mrizka} the Bloch--Floquet decomposition gives four additional conditions, namely
\begin{equation}\label{Floquet}
\eqalign{
\psi_1(\ell/2)=\e^{\im\theta_1}\psi_3(-\ell/2)\,,&\qquad\quad \psi_2(\ell/2)=\e^{\im\theta_2}\psi_4(-\ell/2)\,,\\
\psi_1'(\ell/2)=\e^{\im\theta_1}\psi_3'(-\ell/2)\,,&\qquad\quad \psi_2'(\ell/2)=\e^{\im\theta_2}\psi_4'(-\ell/2)\,,
}
\end{equation}
for some quasimomenta values $\theta_1,\theta_2\in(-\pi,\pi]$. Substituting~(\ref{vlnfce}) into~(\ref{Floquet}) enables us to express $C_1^{\pm}$ and $C_2^{\pm}$ in terms of $C_3^{\pm}$ and $C_4^{\pm}$ as follows,
\begin{equation}\label{elim}
\eqalign{
C_1^+&=C_3^+\cdot\e^{\im(\theta_1-k\ell)}\,;\\
C_1^-&=C_3^-\cdot\e^{\im(\theta_1+k\ell)}\,;\\
C_2^+&=C_4^+\cdot\e^{\im(\theta_2-k\ell)}\,;\\
C_2^-&=C_4^-\cdot\e^{\im(\theta_2+k\ell)}\,.
}
\end{equation}
In the next step we use~(\ref{elim}) to eliminate $C_1^{\pm}$ and $C_2^{\pm}$ from~(\ref{vlnfce}); then we substitute~(\ref{vlnfce}) into~(\ref{vazba}). After simple manipulations we arrive at the condition
\begin{equation}\label{matice}
\left[(U(t)-I)M-k(U(t)+I)N\right]\left(\begin{array}{c}C_3^+\\C_3^-\\C_4^+\\C_4^-\end{array}\right)=0\,,
\end{equation}
where the matrices $M$, $N$ are given by
\begin{equation*}
M=\left(\begin{array}{cccc}
\e^{\im(\theta_1-k\ell)}&\e^{\im(\theta_1+k\ell)}&0&0\\
0&0&e^{\im(\theta_2-k\ell)}&\e^{\im(\theta_2+k\ell)}\\
1&1&0&0\\
0&0&1&1
\end{array}
\right)\,,
\end{equation*}
\begin{equation*}
N=\left(\begin{array}{cccc}
\e^{\im(\theta_1-k\ell)}&-\e^{\im(\theta_1+k\ell)}&0&0\\
0&0&\e^{\im(\theta_2-k\ell)}&-\e^{\im(\theta_2+k\ell)}\\
-1&1&0&0\\
0&0&-1&1
\end{array}
\right)\,.
\end{equation*}
It follows from~(\ref{elim}) that the functions~(\ref{vlnfce}) correspond to a nonzero solution \emph{iff} $\left(C_3^+,C_3^-,C_4^+,C_4^-\right)$ is a nonzero vector. Consequently, a number $k^2$ belongs to the spectrum if and only if there exists a pair $(\theta_1,\theta_2)$ such that (\ref{matice}) has a nontrivial solution, which is equivalent to
\begin{equation*}
\det\left[(U(t)-I)M-k(U(t)+I)N\right]=0\,.
\end{equation*}
Using formula (\ref{Fourier}), one can transform the spectral condition into
\begin{equation}\label{ObecSp}
\det\left[(D-I)F^*M-k(D+I)F^*N\right]=0\,,
\end{equation}
which is easier to deal with. The determinant at the left hand side of (\ref{ObecSp}) equals
\begin{equation*}
512\,\e^{\im(\theta_1+\theta_2)}\,\e^{-\im\frac{(1-t)\gamma}{2}}\left[V_3k^3+V_2k^2+V_1k+V_0\right],
\end{equation*}
where
\begin{equation}\label{V}
\eqalign{
V_3&=-\cos\frac{(1-t)\gamma}{2}\sin^2\frac{\pi t}{4}\sin k\ell(\cos\theta_1+\cos\theta_2+2\cos k\ell)\,; \\
V_2&=2\sin\frac{(1-t)\gamma}{2}\sin^2\frac{\pi t}{4}(\cos\theta_1+\cos k\ell)(\cos\theta_2+\cos k\ell)\,; \\
V_1&=\cos\frac{(1-t)\gamma}{2}\cos^2\frac{\pi t}{4}\sin k\ell(\cos\theta_1+\cos\theta_2-2\cos k\ell)\,; \\
V_0&=-2\sin\frac{(1-t)\gamma}{2}\cos^2\frac{\pi t}{4}\sin^2 k\ell\,.
}
\end{equation}
Therefore, $k^2$ belongs to the spectrum if and only if there exist $\theta_1,\theta_2\in[-\pi,\pi)$ such that
\begin{equation}\label{SpP}
V_3k^3+V_2k^2+V_1k+V_0=0\,.
\end{equation}
In what follows we are going to investigate solutions to this equation.
We will begin with the particular case $\alpha=0$ in Section~\ref{Sect.alpha=0}; the general case will be treated in Section~\ref{Sect.alpha gen.}. Numerical results and observed spectral effects will be discussed in Section~\ref{Sect.numerical}.

\section{Spectrum of a square lattice: Case $\alpha=0$}\label{Sect.alpha=0}

If $\alpha=0$, the $\delta$ coupling is reduced to the Kirchhoff one. In this situation $\gamma=\arg\frac{n+\im\alpha}{n-\im\alpha}=0$, cf.~(\ref{gamma}), and therefore the coefficients $V_2$ and $V_0$ given in~(\ref{V}) vanish. According to~(\ref{SpP}), a number $k^2>0$ belongs to the spectrum if there are $\theta_1,\theta_2\in[-\pi,\pi)$ such that
\begin{eqnarray*}
&& -\sin^2\frac{\pi t}{4}\sin k\ell\,(\cos\theta_1+\cos\theta_2+2\cos k\ell)\, k^3
\\ &&\;\;+\cos^2\frac{\pi t}{4}\sin k\ell\,(\cos\theta_1+\cos\theta_2-2\cos k\ell)\,k=0\,,
\end{eqnarray*}
i.e.,
$$
\sin k\ell\left[\left(k^2\sin^2\frac{\pi t}{4}-\cos^2\frac{\pi t}{4}\right)(\cos\theta_1+\cos\theta_2)-2\left(k^2\sin^2\frac{\pi t}{4}+\cos^2\frac{\pi t}{4}\right)\cos k\ell\right]=0\,.
$$
The spectral condition has two types of solutions. The first are the values of $k$ satisfying $\sin k\ell=0$ giving rise to infinitely degenerate eigenvalues, usually dubbed `Dirichlet', in the positive spectrum. Explicitly, they are squares of
\begin{equation*}
\frac{m\pi}{\ell} \quad (\mbox{for } m\in\mathbb{N})\,.
\end{equation*}
The corresponding eigenfunctions can be supported on finite loops of the lattice. Note that these eigenvalues are independent of the interpolation parameter $t$.

The other type of solutions corresponds to values $k$ such that
$$
\left(k^2\sin^2\frac{\pi t}{4}-\cos^2\frac{\pi t}{4}\right)(\cos\theta_1+\cos\theta_2)=2\left(k^2\sin^2\frac{\pi t}{4}+\cos^2\frac{\pi t}{4}\right)\cos k\ell
$$
for some $\theta_1,\theta_2\in[-\pi,\pi)$. This condition describes spectral bands.
Since $|\cos\theta_j|\leq1,\, j=1,2\,$, one can easily eliminate the term $\cos\theta_1+\cos\theta_2$, obtaining the inequality
\begin{equation}\label{alpha=0}
\left|k^2\sin^2\frac{\pi t}{4}-\cos^2\frac{\pi t}{4}\right|\geq\left(k^2\sin^2\frac{\pi t}{4}+\cos^2\frac{\pi t}{4}\right)|\cos k\ell|\,.
\end{equation}
Note that if $t=0$ (the pure Kirchhoff coupling), the band condition (\ref{alpha=0}) is simplified to $|\cos k\ell|\leq1$, which is satisfied for all $k\ge 0$. Consequently, in the Kirchhoff case the spectrum covers the whole nonnegative part of the real axis. Putting this trivial case aside, from now on we assume that $t\in(0,1]$.

\begin{figure}[h]
\begin{center}
\includegraphics[clip, trim=1cm 2cm 4cm 13.5cm,width=.8\textwidth]{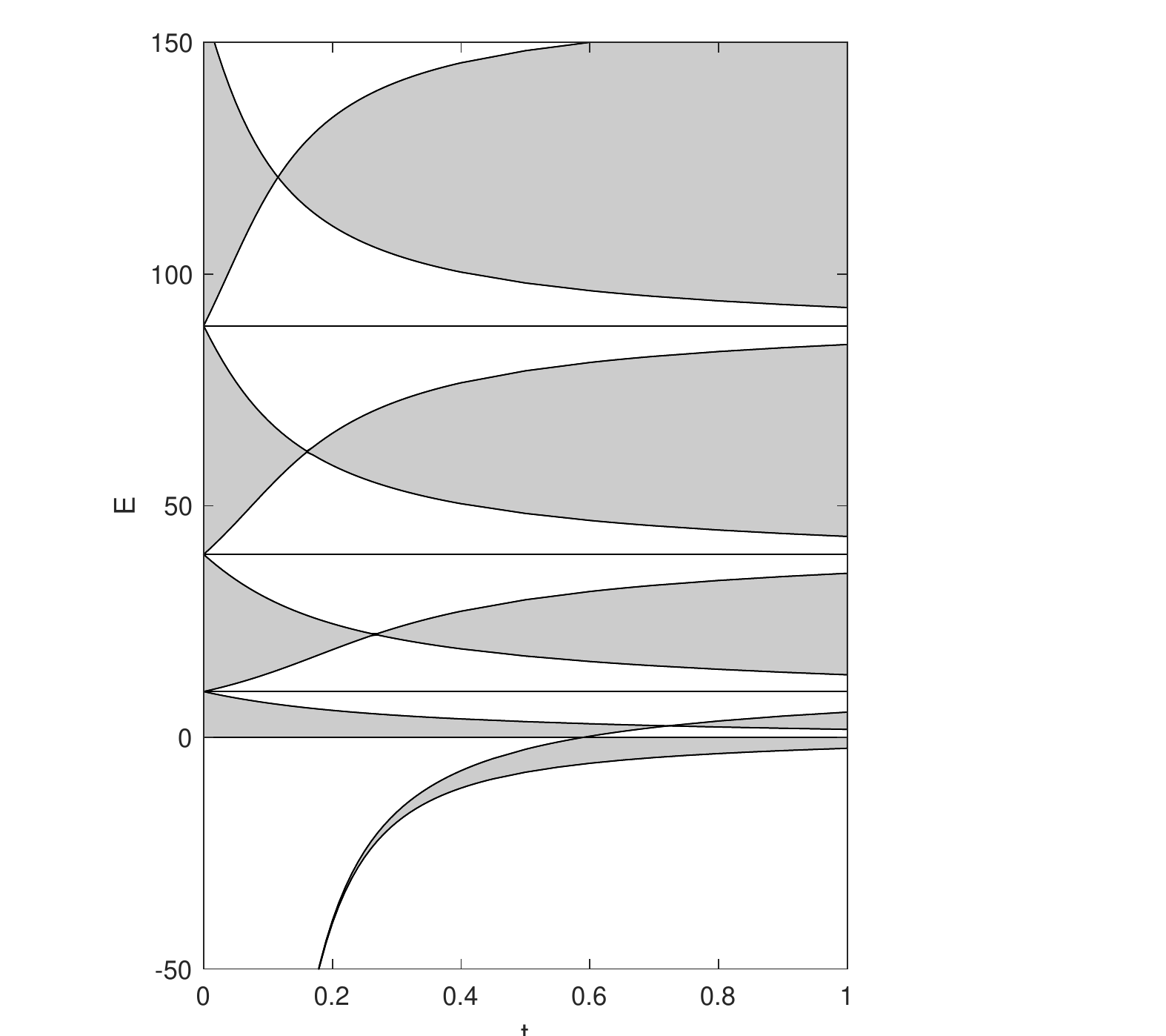}
\caption{Interpolation with the Kirchhoff coupling for $\ell=1$, the spectrum is indicated by the shaded regions.}\label{fig:kirchhoff}
\end{center}
\end{figure}

The band condition (\ref{alpha=0}) is trivially satisfied for $k=(m-1/2)\frac{\pi}{\ell}$ ($m\in\mathbb{N}$) and violated for $k=\frac{m\pi}{\ell}$ ($m\in\mathbb{N}$). Therefore, the spectrum contains infinitely many gaps, which are located in neighborhoods of the points $k=(m\pi/\ell)^2$ for $m\in\mathbb{N}$. In order to describe the bands in a more explicit way, we transform (\ref{alpha=0}) using the equivalence $|a|\geq|b|\Leftrightarrow(a+b)(a-b)\geq0$ and simple trigonometric identities into the form
\begin{equation}\label{band condition}
\sin^2 k\ell\left(k^2\tan^2\frac{k\ell}{2}-\cot^2\frac{\pi t}{4}\right)\left(k^2\cot^2\frac{k\ell}{2}-\cot^2\frac{\pi t}{4}\right)\geq0\,.
\end{equation}
Hence, if $k>0$ (positive spectrum), the band condition~(\ref{band condition}) reads
\begin{equation}\label{band 0+}
\left(k\left|\tan\frac{k\ell}{2}\right|-\cot\frac{\pi t}{4}\right)\left(k\left|\cot\frac{k\ell}{2}\right|-\cot\frac{\pi t}{4}\right)\geq0\,;
\end{equation}
i.e., the spectral bands are located between the curves
\begin{equation*}
k\left|\tan\frac{k\ell}{2}\right|=\cot\frac{\pi t}{4} \quad\mbox{and}\quad k\left|\cot\frac{k\ell}{2}\right|=\cot\frac{\pi t}{4}\,
\end{equation*}
with the points $k=\left(m-\frac{1}{2}\right)\frac{\pi}{\ell}$ \,(for $m\in\mathbb{N}$) inside the bands. It is important to notice that these curves intersect at points where $k\left|\tan\frac{k\ell}{2}\right|=k\left|\cot\frac{k\ell}{2}\right|=\cot\frac{\pi t}{4}$, i.e., at $(t,k)$ such that
\begin{equation}\label{intersect}
k=\left(m-\frac{1}{2}\right)\frac{\pi}{\ell} \quad\mbox{and}\quad t=\frac{4}{\pi}\arccot\left(m-\frac{1}{2}\right)\frac{\pi}{\ell}\,, \quad m\in\mathbb{N}\,.
\end{equation}
This means that at the indicated values of $t$ the spectrum shrinks into a `flat band'. Note that such a parameter-dependent degeneration of spectral bands in periodic quantum graphs is also known in other contexts \cite{EV17}, the difference is that here each band degenerates at a different value of $t$. The situation is illustrated in Figure~\ref{fig:kirchhoff}, which shows how the spectrum depends on $t$ for the lattice spacing $\ell=1$.

Let us look at the asymptotic behaviour of the gaps at high energies, $k\to\infty$. For that purpose it is convenient to rewrite the \emph{gap condition}
\begin{equation*}
\left|k^2\sin^2\frac{\pi t}{4}-\cos^2\frac{\pi t}{4}\right|<\left(k^2\sin^2\frac{\pi t}{4}+\cos^2\frac{\pi t}{4}\right)|\cos k\ell|
\end{equation*}
for large $k$ as follows,
\begin{equation*}
\left|\left(1+\frac{2\cos^2\frac{\pi t}{4}}{k^2\sin^2\frac{\pi t}{4}-\cos^2\frac{\pi t}{4}}\right)\cos k\ell\right|>1\,,
\end{equation*}
which yields, in particular,
\begin{equation}\label{gap 0}
\left|\left(1+\frac{2}{k^2}\cot^2\frac{\pi t}{4}+\mathcal{O}(k^{-4})\right)\cos k\ell\right|>1 \quad \mbox{as}\quad k\to\infty\,.
\end{equation}
We observe that in the high-energy asymptotic regime condition~(\ref{gap 0}) can be satisfied only for $k$ located in small neighborhoods of the points $\frac{m\pi}{\ell}$; namely, large values of $k$ solve (\ref{gap 0}) \emph{iff}
\begin{equation*}
\frac{m\pi}{\ell}-\frac{2}{m\pi}\cot\frac{\pi t}{4}+\mathcal{O}(m^{-2})<k<\frac{m\pi}{\ell}+\frac{2}{m\pi}\cot\frac{\pi t}{4}+\mathcal{O}(m^{-2})
\end{equation*}
for $m\in\mathbb{N}$. The approximate width of the $m$-th spectral gap is thus equal to
\begin{equation*}
\frac{4}{m\pi}\cot\frac{\pi t}{4}+\mathcal{O}(m^{-2})
\end{equation*}
in the momentum variable. It means that the gaps are asymptotically constant in energy, having the widths $\frac{8}{\ell}\cot\frac{\pi t}{4}+\mathcal{O}(m^{-1})$.

\begin{remark}\label{Rem.t=1}
This conclusion corresponds with the asymptotic width value $\approx\frac{8}{\ell}$ at $t=1$ known from \cite{ETa18}. The leading term of the width expansion obtained above also shows that the gap width increases as $t$ diminishes from this extreme value. This is in no contradiction with the fact that the $\delta$ coupling produces no gaps for $\alpha=0$; it is clear that from the intersections specified by (\ref{intersect}) the widths decrease to the point of vanishing at $t=0$.
\end{remark}

It remains to discuss the negative part of the spectrum. The negative spectral condition is obtained from the band condition~(\ref{band condition}) by replacing $k$ with $\im\kappa$ for $\kappa>0$. This gives
\begin{equation*}
-\sinh^2 \kappa\ell\left(\kappa^2\tanh^2\frac{\kappa\ell}{2}-\cot^2\frac{\pi t}{4}\right)\left(\kappa^2\coth^2\frac{\kappa\ell}{2}-\cot^2\frac{\pi t}{4}\right)\geq0\,,
\end{equation*}
which is equivalent to
\begin{equation}\label{band 0-}
\left(\kappa\tanh\frac{\kappa\ell}{2}-\cot\frac{\pi t}{4}\right)\left(\kappa\coth\frac{\kappa\ell}{2}-\cot\frac{\pi t}{4}\right)\leq0\,.
\end{equation}
Points of the negative spectrum are thus determined by the inequalities
\begin{equation}\label{negband}
\kappa\tanh\frac{\kappa\ell}{2}\leq\cot\frac{\pi t}{4}\leq\kappa\coth\frac{\kappa\ell}{2}\,.
\end{equation}
Consequently, there is a single negative spectral band for each $t\in(0,1]$. In particular, if $t\to0$, we have $\cot(\pi t/4)\to\infty$, i.e., the band moves towards large negative energies. Furthermore, since $\tanh(\kappa\ell/2)\approx1$ and $\coth(\kappa\ell/2)\approx1$ for large $\kappa$, we deduce that the negative band for sufficiently small $t$ is located in a neighbourhood of the value $\kappa^2$ for $\kappa=\cot(\pi t/4)$ as can be again seen from Figure~\ref{fig:kirchhoff}. As $t$ approaches $0$, the band shrinks and its width converges to $0$.

Finally, let us examine the spectrum in the vicinity of zero. The positive band condition~(\ref{band 0+}) for $k\to0$ simplifies to
\begin{equation}\label{posband0}
\frac{2}{\ell}\leq\cot\frac{\pi t}{4}\,,
\end{equation}
while the negative band condition~(\ref{negband}) for $\kappa\to0$ becomes
\begin{equation}\label{negband0}
\frac{2}{\ell}\geq\cot\frac{\pi t}{4}\,.
\end{equation}
From here we infer that there are three possibilities of spectral behaviour around zero:
\begin{itemize}
\item If $t<\frac{4}{\pi}\arctan\frac{\ell}{2}$, condition~(\ref{posband0}) is satisfied, while (\ref{negband0}) is not. Therefore, there is a band in the right neighbourhood of $0$, which ends at the value zero, and a gap in the left neighbourhood of zero.
\item If $t>\frac{4}{\pi}\arctan\frac{\ell}{2}$, condition~(\ref{posband0}) is violated, while (\ref{negband0}) is satisfied. Hence, there is a band in the left neighbourhood of $0$, which ends at the value zero, and a gap in the right neighbourhood of zero.
\item If $t=\frac{4}{\pi}\arctan\frac{\ell}{2}$, both the positive spectral band condition~(\ref{band 0+}) and the negative band condition~(\ref{negband}) are satisfied for small $k>0$ and small $\kappa>0$, respectively. In this case zero is located inside a spectral band.
\end{itemize}
This effect is illustrated by Figure~\ref{fig:kirchhoff}. The graph is plotted for $\ell=1$; therefore, the band around zero appears at the value $t=\frac{4}{\pi}\arctan\frac{1}{2}\approx0.5903$.

\section{Spectrum of a square lattice: General $\alpha\in\mathbb{R}$}\label{Sect.alpha gen.}

\subsection{Positive spectrum}\label{Subs. positive}

If $\alpha\neq0$, the spectrum acquires more complexity, but as we will see, it still partly preserves the structure revealed for $\alpha=0$. However, before we proceed to the spectral analysis of a general case, let us single out two special values of $t$ that correspond to the known situations:
\begin{itemize}
\item With reference to the analysis of the extremal case $t=1$ carried out in \cite{ETa18}, see also Remark~\ref{Rem.t=1}, we may assume here and in the following that $t<1$.
\item Also the case $t=0$ corresponding to the pure $\delta$ coupling with a parameter $\alpha\neq0$ is well understood \cite{ET10}. Equation~(\ref{SpP}) together with (\ref{V}) lead to the condition
\begin{equation}\label{band t=0}
\left|\cos k\ell+\frac{1}{k}\tan\frac{\gamma}{2}\sin k\ell\right|\leq1\,.
\end{equation}
The spectrum thus has a band-gap structure. Since $n=4$, equation~(\ref{gamma}) implies $\tan\frac{\gamma}{2}=\frac{\alpha}{4}$ and~(\ref{band t=0}) coincides up to the factor two in the coupling constant with that of the Kronig-Penney model \cite[Sec.~II.2.3]{AGHH}. In particular, $k^2>0$ belongs to a high-energy gap if the following asymptotical conditions in terms of the gap number $m\in\mathbb{N}$ are satisfied,
\begin{eqnarray*}
\frac{m\pi}{\ell}+\mathcal{O}(m^{-2})<k<\frac{m\pi}{\ell}+\frac{2}{m\pi}\tan\frac{\gamma}{2}+\mathcal{O}(m^{-2}) \qquad \mbox{for } \gamma>0\,; \\
\frac{m\pi}{\ell}+\frac{2}{m\pi}\tan\frac{\gamma}{2}+\mathcal{O}(m^{-2})<k<\frac{m\pi}{\ell}+\mathcal{O}(m^{-2}) \qquad \mbox{for } \gamma<0\,.
\end{eqnarray*}
\end{itemize}

From now on we thus consider $t\in(0,1)$ only. We start the analysis from the spectral condition~(\ref{SpP}). With regard to the structure of expressions $V_j$ and using the assumptions $\gamma\neq0$, $t\neq0$, and $t\neq1$, we rewrite it as
\begin{equation}\label{ABC}
\cos\theta_1\cos\theta_2+A(\cos\theta_1+\cos\theta_2)+B=0\,,
\end{equation}
where
\begin{equation}\label{A..}
A=-\frac{1}{2}\cot\frac{(1-t)\gamma}{2}\left(k-\frac{1}{k}\cot^2\frac{\pi t}{4}\right)\sin k\ell+\cos k\ell
\end{equation}
and
\begin{equation}\label{B..}
\hspace{-5em} B=-\cot\frac{(1-t)\gamma}{2}\left(k+\frac{1}{k}\cot^2\frac{\pi t}{4}\right)\cos k\ell\sin k\ell+\cos^2 k\ell-\frac{1}{k^2}\cot^2\frac{\pi t}{4}\sin^2 k\ell\,.
\end{equation}
Rewriting~(\ref{ABC}) in the form
\begin{equation*}
(\cos\theta_1+A)(\cos\theta_2+A)=A^2-B,
\end{equation*}
we reformulate the spectral condition as follows,
\begin{equation}\label{sp.c.}
\hspace{-6em} \min_{\theta_1,\theta_2\in[-\pi,\pi)}(\cos\theta_1+A)(\cos\theta_2+A)\leq A^2-B \leq \max_{\theta_1,\theta_2\in[-\pi,\pi)}(\cos\theta_1+A)(\cos\theta_2+A)\,.
\end{equation}
We observe that
\begin{equation*}
\max_{\theta_1,\theta_2\in[-\pi,\pi)}(\cos\theta_1+A)(\cos\theta_2+A)=(1+|A|)^2\,,
\end{equation*}
$$
\min_{\theta_1,\theta_2\in[-\pi,\pi)}(\cos\theta_1+A)(\cos\theta_2+A)=
\left\{\begin{array}{ll}
(-1+|A|)^2 & \mbox{for } 1\leq|A|\,; \\
(-1+|A|)(1+|A|) & \mbox{for } 1>|A|\,.
\end{array}\right.
$$
This allows us to eliminate the Bloch variables $\cos\theta_1,\cos\theta_2$ from (\ref{sp.c.}). After a simple manipulation, we obtain the spectral condition in the following form:
\begin{itemize}
\item either $|A|\geq1$ and
\begin{equation*}
|B+1|\leq2|A|\,,
\end{equation*}
\item or $|A|<1$ and
\begin{equation*}
|B+|A||\leq 1+|A|\,,
\end{equation*}
\end{itemize}
where $A,B$ are given by~(\ref{A..}) and (\ref{B..}), respectively.
The system of inequalities
$$
(|A|\geq1\;\wedge\;|B+1|\leq2|A|) \quad\vee\quad (|A|<1\;\wedge\;|B+|A||\leq 1+|A|)
$$
can be equivalently expressed in the following manner,
\begin{equation}\label{sp podm}
\fl (B+1-2A)(B+1+2A)\leq0 \quad\vee\quad (B+1-2A\geq0 \wedge B+1+2A\geq0 \wedge B\geq1)\,,
\end{equation}
which is considerably more convenient, because all the terms occurring in~(\ref{sp podm}) can be easily factorized:
$$
B+1-2A=-\frac{\sin k\ell}{k}(1+\cos k\ell)\left(\cot\frac{(1-t)\gamma}{2}+\frac{1}{k}\tan\frac{k\ell}{2}\right)\left(\cot^2\frac{\pi t}{4}-k^2\tan^2\frac{k\ell}{2}\right);
$$
$$
B+1+2A=\frac{\sin k\ell}{k}(1-\cos k\ell)\left(\cot\frac{(1-t)\gamma}{2}-\frac{1}{k}\cot\frac{k\ell}{2}\right)\left(\cot^2\frac{\pi t}{4}-k^2\cot^2\frac{k\ell}{2}\right);
$$
$$
B-1=-\sin^2 k\ell\left(k\cot k\ell\cot\frac{(1-t)\gamma}{2}+1\right)\left(\frac{1}{k^2}\cot^2\frac{\pi t}{4}+1\right).
$$
Using the expressions above, we can rewrite the spectral condition~(\ref{sp podm}) in the form of products of relatively simple factors, namely:
\begin{itemize}
\item either
\begin{equation}\label{sp podm 1}
\eqalign{
\frac{1}{k^2}\left(\cot\frac{(1-t)\gamma}{2}+\frac{1}{k}\tan\frac{k\ell}{2}\right)
\left(\cot\frac{(1-t)\gamma}{2}-\frac{1}{k}\cot\frac{k\ell}{2}\right) \\
\times \left(\cot^2\frac{\pi t}{4}-k^2\tan^2\frac{k\ell}{2}\right)
\left(\cot^2\frac{\pi t}{4}-k^2\cot^2\frac{k\ell}{2}\right)\geq0\,,
}
\end{equation}
\item or
\begin{equation}\label{sp podm 2}
\hspace{-2cm}
\left.\eqalign{
\frac{\sin k\ell}{k}(1+\cos k\ell)\left(\cot\frac{(1-t)\gamma}{2}+\frac{1}{k}\tan\frac{k\ell}{2}\right)\left(\cot^2\frac{\pi t}{4}-k^2\tan^2\frac{k\ell}{2}\right)\leq0 \\
\frac{\sin k\ell}{k}(1-\cos k\ell)\left(\cot\frac{(1-t)\gamma}{2}-\frac{1}{k}\cot\frac{k\ell}{2}\right)\left(\cot^2\frac{\pi t}{4}-k^2\cot^2\frac{k\ell}{2}\right)\geq0 \\
\sin^2 k\ell\left(k\cot k\ell\cot\frac{(1-t)\gamma}{2}+1\right)\left(\frac{1}{k^2}\cot^2\frac{\pi t}{4}+1\right)\geq0\,.
}\right\}
\end{equation}
\end{itemize}
This spectral condition is meant as an alternative, (\ref{sp podm 1})~$\vee$~(\ref{sp podm 2}), where~(\ref{sp podm 2}) requires all the three involved inequalities to be satisfied.
The factors occurring in~(\ref{sp podm 1}) and (\ref{sp podm 2}) correspond to the boundaries of the spectral bands.
For example, all the pairs $(t,k)$ obeying the condition~(\ref{sp podm 1}) for $k>0$ take the form of a union of certain areas bounded by (some subset of) the curves
\begin{eqnarray*}
&& \cot\frac{(1-t)\gamma}{2}+\frac{1}{k}\tan\frac{k\ell}{2}=0\,,\qquad
\cot\frac{(1-t)\gamma}{2}-\frac{1}{k}\cot\frac{k\ell}{2}=0\,, \\ &&
\cot^2\frac{\pi t}{4}-k^2\tan^2\frac{k\ell}{2}=0\,,\qquad
\cot^2\frac{\pi t}{4}-k^2\cot^2\frac{k\ell}{2}=0\,.
\end{eqnarray*}
Note that the latter two curves are independent of $\gamma$: in other words, they are independent of the coupling constant $\alpha$, and they coincide with the analogous curves that appeared in Section~\ref{Sect.alpha=0}. As a consequence, the endpoints of spectral bands for $\alpha\neq0$ partly coincide with the endpoints of the bands of the system with Kirchhoff couplings ($\alpha=0$).

\begin{figure}[h]
\begin{center}
\includegraphics[clip, trim=1cm 2cm 4cm 14cm,width=.8\textwidth]{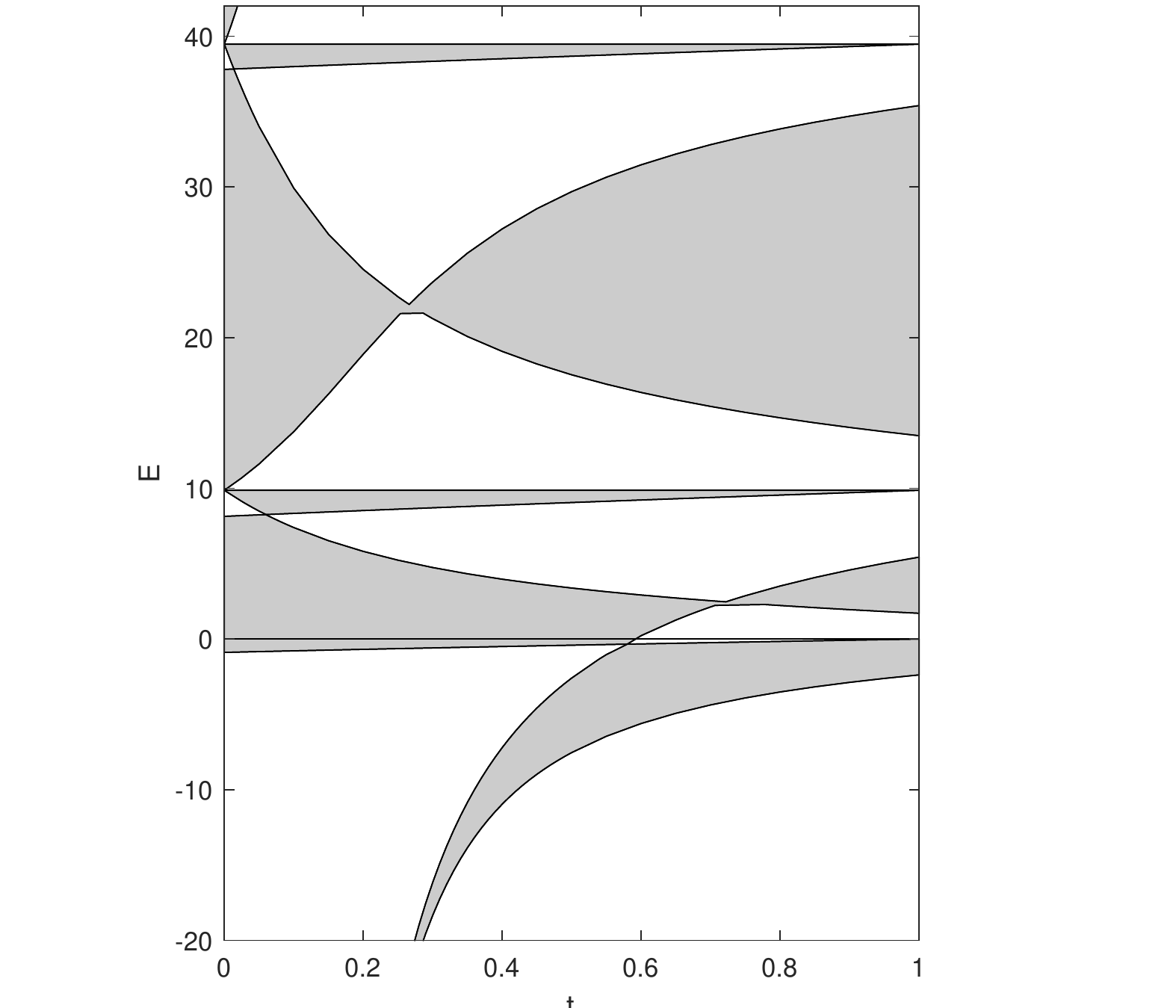}
\caption{Interpolation with an attractive $\delta$ coupling, $\alpha=-4(\sqrt{2}-1)$, $\ell=1$.}\label{fig:negative}
\end{center}
\end{figure}

\begin{figure}[h]
\begin{center}
\includegraphics[clip, trim=1cm 2cm 4cm 14cm,width=.8\textwidth]{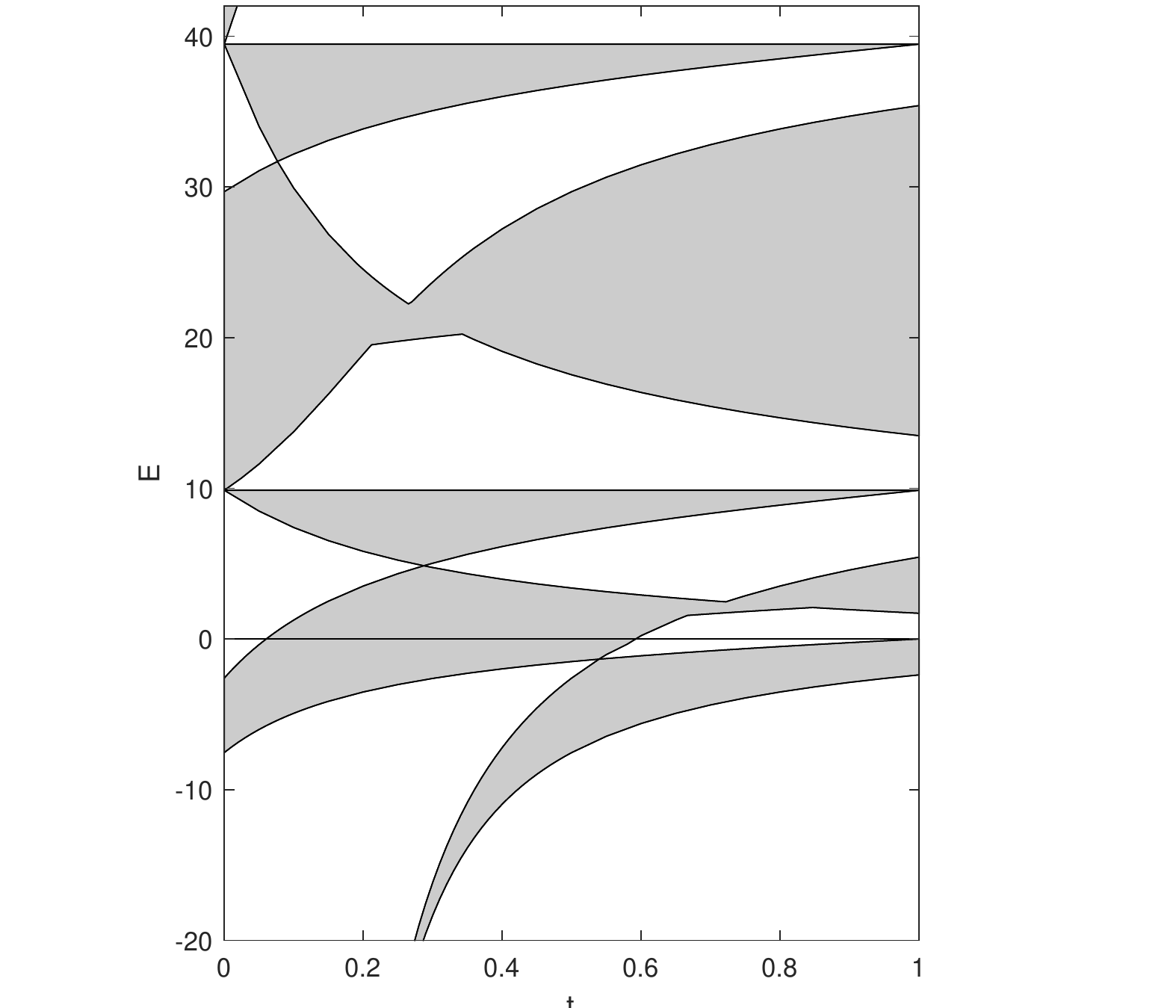}
\caption{Interpolation with an attractive $\delta$ coupling, $\alpha=-4(\sqrt{2}+1)$, $\ell=1$.}\label{fig:negative2}
\end{center}
\end{figure}

In Figures~\ref{fig:negative} and~\ref{fig:negative2} we plot the spectra for the lattice spacing $\ell=1$ and the cases of a weakly and strongly attractive $\delta$ coupling, $\alpha=-4(\sqrt{2}\mp 1)$ corresponding to $\gamma=-\im\pi/4$ and $\gamma=-3\im\pi/4$, in other words, $\alpha\approx -1.65685$ and $\alpha\approx -9.65685$, respectively. The plots include the negative spectrum which we are going to discuss in the next section.

The conditions~(\ref{sp podm 1}) and (\ref{sp podm 2}) give rise to various spectral band regimes, depending on the parameters $\ell$ and $\gamma$. We will return to that and discuss some of them in Section~\ref{Sect.numerical}.

\subsection{Negative spectrum}

The negative spectral condition is obtained straightforwardly from the condition derived in Section~\ref{Subs. positive} by replacing $k$ with $\im\kappa$ for $\kappa>0$. Therefore, a number $-\kappa^2<0$ belongs to the negative spectrum if
\begin{itemize}
\item either
\begin{eqnarray*}
&& \frac{1}{-\kappa^2}\left(\cot\frac{(1-t)\gamma}{2}+\frac{1}{\kappa}\tanh\frac{\kappa\ell}{2}\right)
\left(\cot\frac{(1-t)\gamma}{2}+\frac{1}{\kappa}\coth\frac{\kappa\ell}{2}\right) \\ &&
\times \left(\cot^2\frac{\pi t}{4}-\kappa^2\tanh^2\frac{\kappa\ell}{2}\right)
\left(\cot^2\frac{\pi t}{4}-\kappa^2\coth^2\frac{\kappa\ell}{2}\right)\geq0\,,
\end{eqnarray*}
\item or
\begin{eqnarray*}
\fl \frac{\sinh \kappa\ell}{\kappa}(1+\cosh \kappa\ell)\left(\cot\frac{(1-t)\gamma}{2}+\frac{1}{\kappa}\tanh\frac{\kappa\ell}{2}\right)\left(\cot^2\frac{\pi t}{4}-\kappa^2\tanh^2\frac{\kappa\ell}{2}\right)\leq0 \\
\fl \wedge\quad
\frac{\sinh \kappa\ell}{\kappa}(1-\cosh \kappa\ell)\left(\cot\frac{(1-t)\gamma}{2}+\frac{1}{\kappa}\coth\frac{\kappa\ell}{2}\right)\left(\cot^2\frac{\pi t}{4}-\kappa^2\coth^2\frac{\kappa\ell}{2}\right)\geq0 \\
\fl \wedge\quad
-\sinh^2 \kappa\ell\left(\kappa\coth\kappa\ell\cot\frac{(1-t)\gamma}{2}+1\right)\left(\frac{1}{-\kappa^2}\cot^2\frac{\pi t}{4}+1\right)\geq0\,.
\end{eqnarray*}
\end{itemize}
This can be simplified to
\begin{itemize}
\item either
\begin{equation}\label{sp podm negat}
\eqalign{
\left(\cot\frac{(1-t)\gamma}{2}+\frac{1}{\kappa}\tanh\frac{\kappa\ell}{2}\right)
\left(\cot\frac{(1-t)\gamma}{2}+\frac{1}{\kappa}\coth\frac{\kappa\ell}{2}\right) \nonumber \\
\times \left(\cot\frac{\pi t}{4}-\kappa\tanh\frac{\kappa\ell}{2}\right)
\left(\cot\frac{\pi t}{4}-\kappa\coth\frac{\kappa\ell}{2}\right)\leq0\,,
}
\end{equation}
\item or
\begin{equation}\label{sp podm negat 2}
\left.\eqalign{
\left(\cot\frac{(1-t)\gamma}{2}+\frac{1}{\kappa}\tanh\frac{\kappa\ell}{2}\right)\left(\cot\frac{\pi t}{4}-\kappa\tanh\frac{\kappa\ell}{2}\right)\leq0 \\
\left(\cot\frac{(1-t)\gamma}{2}+\frac{1}{\kappa}\coth\frac{\kappa\ell}{2}\right)\left(\cot\frac{\pi t}{4}-\kappa\coth\frac{\kappa\ell}{2}\right)\leq0 \\
\left(\cot\frac{(1-t)\gamma}{2}+\frac{1}{\kappa}\tanh\kappa\ell\right)\left(\cot\frac{\pi t}{4}-\kappa\right)\geq0\,.
}\right\}
\end{equation}
\end{itemize}
Note that if $\gamma>0$, in other words, $\alpha>0$, every factor in~(\ref{sp podm negat}) and~(\ref{sp podm negat 2}) that depends on $\gamma$ is positive. Then~(\ref{sp podm negat 2}) has no solution, because the inequalities
$$
\cot\frac{\pi t}{4}-\kappa\tanh\frac{\kappa\ell}{2}\leq0\,,\quad \cot\frac{\pi t}{4}-\kappa\geq0
$$
cannot be satisfied at the same time due to the fact that $\tanh(\kappa\ell/2)<1$. The spectral condition is thus given only by~(\ref{sp podm negat}), which simplifies to
\begin{equation*}
\left(\cot\frac{\pi t}{4}-\kappa\tanh\frac{\kappa\ell}{2}\right)\left(\cot\frac{\pi t}{4}-\kappa\coth\frac{\kappa\ell}{2}\right)\leq0\,.
\end{equation*}
Comparing this result with~(\ref{band 0-}) we conclude that for a repulsive $\delta$ interaction, meaning $\alpha>0$, the negative spectrum coincides with the negative spectrum for $\alpha=0$.

If $\gamma<0$, i.e. the $\delta$ interaction is attractive, meaning $\alpha<0$, the situation is more intriguing. The system of conditions (\ref{sp podm negat})~$\vee$~(\ref{sp podm negat 2}) implies that a $-\kappa^2<0$ belongs to the spectrum if one of the following conditions is satisfied:
\begin{eqnarray}
&&\!\!\!\!\!\!\!\!\!\!\left.\eqalign{
\kappa\tanh\frac{\kappa\ell}{2}\leq\cot\frac{\pi t}{4}\leq\kappa\coth\frac{\kappa\ell}{2} \\
\quad\wedge\quad
\left(-\cot\frac{(1-t)\gamma}{2}\leq\frac{1}{\kappa}\tanh\frac{\kappa\ell}{2} \;\vee\; -\cot\frac{(1-t)\gamma}{2}\geq\frac{1}{\kappa}\coth\frac{\kappa\ell}{2}\right);
}\right\} \label{neg band 1} \\
&&\!\!\!\!\!\!\!\!\!\!\left.\eqalign{
\frac{1}{\kappa}\tanh\frac{\kappa\ell}{2}\leq-\cot\frac{(1-t)\gamma}{2}\leq\frac{1}{\kappa}\coth\frac{\kappa\ell}{2} \\
\quad\wedge\quad
\left(\cot\frac{\pi t}{4}\leq\kappa\tanh\frac{\kappa\ell}{2} \;\vee\; \cot\frac{\pi t}{4}\geq\kappa\coth\frac{\kappa\ell}{2}\right);
}\right\} \label{neg band 2} \\
&&\!\!\!\!\!\!\!\!\!\! \kappa\tanh\frac{\kappa\ell}{2}\leq\cot\frac{\pi t}{4}\leq\kappa \quad\wedge\quad
\frac{1}{\kappa}\tanh\kappa\ell\leq-\cot\frac{(1-t)\gamma}{2}\leq\frac{1}{\kappa}\coth\frac{\kappa\ell}{2}\,; \label{neg band 3} \\
&&\!\!\!\!\!\!\!\!\!\! \kappa\leq\cot\frac{\pi t}{4}\leq\kappa\coth\frac{\kappa\ell}{2} \quad\wedge\quad
\frac{1}{\kappa}\tanh\frac{\kappa\ell}{2}\leq-\cot\frac{(1-t)\gamma}{2}\leq\frac{1}{\kappa}\tanh\kappa\ell\,. \label{neg band 4}
\end{eqnarray}
The solution of~(\ref{neg band 1}) constitutes a band that coincides at sufficiently large negative energies with the negative band found already in Section~\ref{Sect.alpha=0} for $\alpha=0$. The solution of~(\ref{neg band 2}) constitutes another band, which is present only for $\alpha<0$.
The solutions of~(\ref{neg band 3}) and~(\ref{neg band 4}), if present, are attached to the areas represented by~(\ref{neg band 1}) and~(\ref{neg band 2}), respectively.

\subsection{The limits $t\to0$ and $t\to1$}\label{Sect.lim}

Let us finally look at the behavior of the spectrum as $t$ approaches the extremes of the interpolation interval, $t=0$ and $t=1$. The endpoints of spectral bands for a given $t\in(0,1)$ lie on the curves $F(t,k)=0$, where $F(t,k)$ are the equations occurring as factors of the spectral conditions~(\ref{sp podm 1}) and (\ref{sp podm 2}). We will demonstrate that in the limit $t\to0$ and $t\to1$, the band edges of the mixed system continuously approach the band edges of the system with the pure $\delta$ coupling and the `extremal' rotationally symmetric coupling, respectively. The obvious exception to this claim is the lowest spectral band which disappears in the limit $t\to0$. We will also show that the band-edge dependence in the vicinity of the extremes is approximately linear in $t$, with a nonzero derivative.

For the sake of brevity, we consider the band edge given implicitly as a solution of the equation $F(t,k)=0$, where
\begin{equation}\label{Ftk}
F(t,k)=\cot\frac{(1-t)\gamma}{2}+\frac{1}{k}\tan\frac{k\ell}{2}\,;
\end{equation}
the analysis of other functions would be similar. We also assume that $\cot\frac{\gamma}{2}$ is finite; otherwise we would equivalently analyze the implicit function $\tilde{F}(t,k)=0$ for $\tilde{F}(t,k)=\tan\frac{(1-t)\gamma}{2}+k\cot\frac{k\ell}{2}$. The derivative of the implicit function $k=k(t)$ is, according to the classical formula,
\begin{equation}\label{Ftk der}
\frac{\d k}{\d t}=-\frac{\frac{\partial F}{\partial t}}{\frac{\partial F}{\partial k}}=-\frac{\frac{-1}{\sin^2\frac{(1-t)\gamma}{2}}\cdot\frac{-\gamma}{2}}{-\frac{1}{k^2}\tan\frac{k\ell}{2}+\frac{1}{k}\cdot\frac{1}{\cos^2\frac{k\ell}{2}}\cdot\frac{\ell}{2}}\,.
\end{equation}
Note that the expression $\partial F/\partial k$ in the denominator at the right-hand side can never attain zero (unless $k=0$), which is easy to check.

Consider first the case $t\to0$. The limit value $\lim_{t\to0}k(t)=k_0$ is a solution of
\begin{equation*}
\cot\frac{\gamma}{2}+\frac{1}{k_0}\tan\frac{k_0\ell}{2}=0\,.
\end{equation*}
In this case, inequality~(\ref{band 0+}) becomes equality, i.e., such a $k_0$ corresponds to a band edge in the system with the pure $\delta$ coupling of parameter $\gamma$. This implies the continuity of the band edge shift as $t\to0$. The derivative~(\ref{Ftk der}) is finite and moreover nonzero for $\gamma\neq0$, hence when $t$ approaches zero, the band edge of the interpolating system approaches the band edge of that with the pure $\delta$ coupling as $k_t=k_0+c_0t+\mathcal{O}(t^2)$ with some $c_0\neq0$.

On the other hand, for $t\to1\:$ (\ref{Ftk}) implies $\lim_{t\to1}k(t)=(2m-1)\pi/\ell$ for some $m\in\mathbb{N}$. Recall that the value $k_1=(2m-1)\pi/\ell$ indeed corresponds to a band edge of the system with the `extremal' rotationally symmetric coupling, cf.~(\ref{band 0+}). Consequently, the band edge shift is continuous as $t\to1$. In the left neighbourhood of $t=1$, we have $\tan(k\ell/2)=-k\cot((1-t)\gamma/2)$, which we substitute to the derivative~(\ref{Ftk der}); hence we get
$$
\frac{\d k}{\d t}
=-\frac{\frac{\gamma}{2}}
{\frac{1}{k}\sin\frac{(1-t)\gamma}{2}\cos\frac{(1-t)\gamma}{2}
+\frac{1}{k}\left(\sin^2\frac{(1-t)\gamma}{2}+k^2\cos^2\frac{(1-t)\gamma}{2}\right)\frac{\ell}{2}}\,,
\;.
$$
and therefore
$$
\lim_{t\to1}\frac{\d k}{\d t}=-\frac{\frac{\gamma}{2}}{0+\frac{1}{k_1}\left(0+k_1^2\cdot1\right)\frac{\ell}{2}}=\frac{-\gamma}{k_1\ell}=\frac{-\gamma}{\frac{(2m-1)\pi}{\ell}\ell}=\frac{-\gamma}{(2m-1)\pi}\;.
$$
We conclude that the band edge continuously approaches the band edge corresponding to the `extremal' rotationally symmetric coupling introduced in (\ref{U1}), with a rate given as $k_t=k_1+c_1t+\mathcal{O}(t^2)$ with $c_1=-\gamma/(k_1\ell)\neq0$.
\begin{remark}
In a similar vein one can consider the dependence of $E=k^2$ on the interpolation parameter $t$. Since $\d E/\d t=(\d k^2/\d k)\cdot(\d k/\d t)=2k(\d k/\d t)$, the one-sided derivatives $\d E/\d t$ at $t=0$ and $t=1$ are finite and nonzero as well.
\end{remark}

\section{Spectrum of a square lattice: Numerics and discussion}\label{Sect.numerical}

In the previous sections we have already used illustrations coming from numerical solution of the spectral condition (\ref{ObecSp}). Note that the numerical treatment can proceed in two equivalent and complementary ways, either to follow the reduction of the spectral condition as we did in the analytical reasoning above or to evaluate the determinant entering (\ref{ObecSp}) for a fixed $k$ and and to inspect the range of its values over the Brillouin zone, it is needless to say that both ways yield the same result.

\begin{figure}[h]
\begin{center}
\includegraphics[clip, trim=1cm 2cm 4cm 13.5cm,width=.8\textwidth]{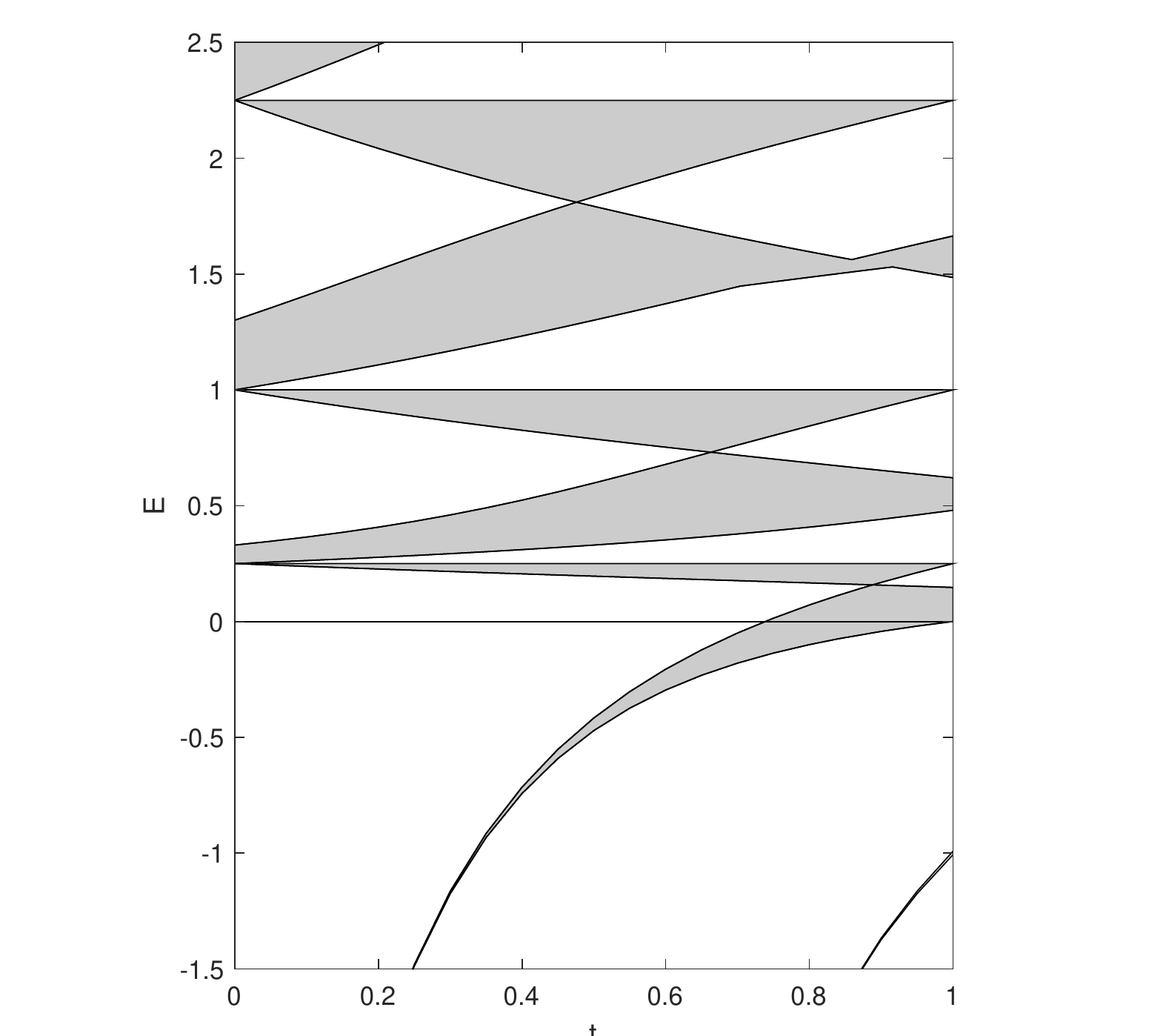}
\caption{Interpolation with an attractive $\delta$ coupling, $\alpha=-4(\sqrt{2}+1)$ and $l=2\pi$.}\label{fig:big_l}
\end{center}
\end{figure}

In the numerical analysis leading to Figures~\ref{fig:kirchhoff}--\ref{fig:negative2} we focused on the Kirchhoff case, $\alpha=0$, and on lattices with attractive $\delta$ interaction; we do not provide illustrations with a repulsive $\delta$ interaction case where the spectrum for $t=0$ is positive, otherwise its behaviour is similar. Our last numerical result, plotted in Figure~\ref{fig:big_l}, shows the dependence of the spectrum on the lattice spacing. We consider the same $\alpha$ as in Figure~\ref{fig:negative2} but now a substantially larger $\ell$. Since  $\ell>2$, the lattice with $t=1$ has, according to \cite{ETa18}, one purely negative spectral band seen in the lower right corner; it disappears to $-\infty$ as $t\to 0$ in accordance with (\ref{negband}). Once $t<1$, there is another component of negative spectrum which evolves into the single negative band which the Kronig-Penney lattice, $t=0$, has in this case (off the scale in the figure).

Let us summarize our observations about the band spectra of the lattice considered:

\begin{itemize}

\item[(i)]\emph{A `discontinuity' at $t=0$:} There is always a spectral band which becomes narrow and strongly negative as the interpolation parameter approaches zero and eventually disappears. This is expected in view of condition~(\ref{negband}) and it corresponds to the behaviour of the `additional' eigenvalues mentioned at the end of Section~\ref{s:star}.

\item[(ii)]\emph{Point degeneracies for $\alpha=0$:} In the Kirchhoff case we have noted that at particular values of $t$ spectral bands may collapse to a point producing a `flat band' or an infinitely degenerate eigenvalue different from the `Dirichlet' eigenvalues. This is in accordance with the theoretical analysis, cf.~(\ref{intersect}). This phenomenon, however, does not apply to the case $\alpha\neq0$ in which both the analytical and numerical results show that the spectral condition~(\ref{sp podm 2}) has other solutions which smear these Kirchhoff degenerate eigenvalues into bands of nonzero width. It is obvious that this smearing becomes more pronounced as we are going farther from the Kirchhoff case.

\item[(iii)]\emph{Non-monotonicity of the gap widths:} As mentioned already in Remark~\ref{Rem.t=1}, in the case $\alpha=0$, the widths of the gaps are not monotonous with respect to the interpolation parameter. The same remains true in the case $\alpha\ne 0$.

\item[(iv)]\emph{$t$-independence of some bands:} As noted at the end of Section~\ref{Subs. positive}, some curves marking the band edges are independent of $\alpha$. This is clearly visible in Figures~\ref{fig:kirchhoff}--\ref{fig:negative2}. Take the band edges of the second positive band in Figure~\ref{fig:kirchhoff} spanning the interval that moves from $[\pi^2,4\pi^2]$ for $t=0$, shrinking to a point $t=\frac{4}{\pi}\arccot\frac{3\pi}{2}\approx 0.266$ and subsequently expanding again as the parameter grows towards $t=1$. One finds the same pair of curves in Figures~\ref{fig:negative} and \ref{fig:negative2}. However, for $\alpha\neq0$ the band edges coincide only in parts of the interval $[0,1]$, because there is a neighbourhood of $t=\frac{4}{\pi}\arccot\frac{3\pi}{2}$ in which the condition~(\ref{sp podm 2}) gives rise to additional solutions, as mentioned in (ii) above. Another difference is that gaps open in the spectrum for $t$ small enough due to intersection with another curve; for larger $t$ the band shape remains unchanged. It is also clear that the $\alpha$-independent curves may in such situations play alternatively the role of an upper or lower band edge for new bands that are not present for $\alpha=0$.

\item[(v)]\emph{Band edge regularity:} The curves delineating the band edges are described by analytic functions. In the Kirchhoff case the analyticity is violated only at the points when the curves are crossing according to (\ref{intersect}). On the other hand, in the case $\alpha\ne 0$ where the spectrum is a union of bands there are other points where each particular edge is not smooth; needless to say, it remains Lipshitz.

\item[(vi)]\emph{Flat band spreading:} Another interesting effect concerns the infinitely degenerate eigenvalues of the `purely rotational' lattice, $t=1$, when the interpolation parameter decreases. On the one hand we know from the above analysis that they remain in the spectrum, on the other hand our analysis shows that they are smeared into a band of increasing width which may finally start shrinking to a point again when its lower edge crosses the upper edge of the band immediately below, cf.~Figures~\ref{fig:negative} and \ref{fig:negative2}. A possible explanation may come from the realization that the `elementary' eigenfunctions in the case $t=1$ are of two types. It was noted in \cite{ETa18} apart from the `Dirichlet' ones (which consist now of sine segments with zeroes at the nodes), there are `Neumann' ones composed of cosine segments with zero derivatives at the nodes; the former may be insensitive to the interpolation, the latter not.

\end{itemize}

\subsection*{Acknowledgments}

The research was supported by the Czech Science Foundation (GA\v{C}R) within the project 17-01706S.


\end{document}